\documentclass[12pt,a4paper]{article}
\usepackage{amsmath,amssymb,amsfonts}
\usepackage{graphicx,epsfig}
\usepackage{hyperref}
\usepackage{color}
\usepackage{amsmath}
\usepackage{amsfonts}
\usepackage{amssymb}
\usepackage{graphicx}
\usepackage{slashed}
\numberwithin{equation}{section}
\newcommand{\be}{\begin{equation}}
\newcommand{\ee}{\end{equation}}
\newcommand{\bea}{\begin{eqnarray}}
\newcommand{\eea}{\end{eqnarray}}
\newcommand{\ba}{\begin{array}}
\newcommand{\ea}{\end{array}}

\newcommand{\cA}{{\cal A}}

\newcommand{\cO}{{\cal O}}
\newcommand{\cH}{{\cal H}}

\newcommand{\no}{\nonumber}

\newcommand{\YU}{Y^u}
\newcommand{\YD}{Y^d}
\newcommand{\YUdag}{Y^{u\dagger}}
\newcommand{\YDdag}{Y^{d\dagger}}
\newcommand{\dis}{\displaystyle}
\newcommand{\Pu}{\hat P_u}
\newcommand{\Pd}{\hat P_d}
\newcommand{\Pud}{\hat P_{u(d)}}

\begin{document}
\begin{titlepage}
\begin{flushright}
\end{flushright}
\vskip 1.0cm
\begin{center}
{\Large \bf Composite fermions in  Electroweak Symmetry Breaking} \\
\vskip 1.0cm
{\large Riccardo Barbieri$^a$,\ Gino Isidori$^{a,b}$,\  
Duccio Pappadopulo$^a$} \\[1cm]
{\it $^a$ Scuola Normale Superiore and INFN, Piazza dei Cavalieri 7, 56126 Pisa, Italy} \\ [2mm]
{\it $^b$ INFN, Laboratori Nazionali di Frascati, Via E. Fermi 40
I-00044 Frascati, Italy}\\
\vskip 2.0cm 
\abstract{If the electroweak symmetry is broken by some unspecified strong dynamics, composite fermions may exist with definite transformation properties under $SU(2)_L\times SU(2)_R/SU(2)_{L+R}$ and may play a role in giving masses by mixing to all the standard quarks and leptons. 
Assuming this to be the case, we analyze the role of Singlets, Doublets and Triplets in the ElectroWeak Precision Tests and in Flavour Physics. 
Doublets and Triplets are generically disfavoured.  In the Singlet case, we specify the breaking patterns of the flavour group that allow to keep the CKM picture of flavour physics and we discuss the effects of the mixing between composite and elementary fermions.  These mixings affect in particular the rather peculiar LHC phenomenology of the composite fermions. }
\end{center}%
\end{titlepage}

\section{Introduction and statement of the framework}

That a perturbative Higgs boson exist, one or more and perhaps supersymmetric, is highly probable, with its elusiveness so far explained by the lack of direct experiments at the Fermi scale or above it. The difficulty of proposing explicit and sensible {\it Higgsless} models,
despite some interesting recent attempts~\cite{Csaki:2003dt,Foadi:2003xa,Georgi:2004iy},
is not the least reason behind this view.
However, while waiting for the LHC to say the final word on this issue, we  find  useful to spend some time in exploring possible generic patterns of Higgsless descriptions  
of ElectroWeak Symmmetry Breaking (EWSB).  
In fact, it is again the very lack of direct experiments at the relevant energy scale that motivates us, since this lack may hide some clues needed to understand the physics of EWSB.

We base our considerations on the following generic picture. Some strong dynamics breaks a $SU(2)_L\times SU(2)_R\times U(1)_X$ symmetry, global in the limit of vanishing electroweak gauge couplings, down to $SU(2)_{L+R}\times U(1)_X$.
This spontaneous symmetry breaking, characterised by the 
scale  $v = (\sqrt{2} G_F)^{-1/2} \approx 246$, 
also leads to the breaking of the 
standard electroweak gauge symmetry, $SU(2)_L\times U(1)_Y,~Y=T_{3R} + X$, 
down to the electromagnetic $U(1)$.
There exist three generations of fermions with the usual $SU(2)_L\times U(1)_Y$ quantum numbers, called {\it elementary} since they do not feel directly the strong dynamics.
The strong dynamics produces composite vectors and composite fermions with definite transformation properties under $SU(2)_L\times SU(2)_R/SU(2)_{L+R}$. These non-linear transformations involve as usual the Goldstone fields, $\hat{\pi} = \pi^a \sigma^a/2$, themselves transforming under $SU(2)_L\times SU(2)_R$ as 
\be
U \rightarrow g_R U g_L^\dagger,~~~U \equiv e^{i 2\hat{\pi}/v}.
\label{U}
\ee
The exchange of the composite vectors is supposed to keep unitary the scattering amplitudes of $W$'s and $Z$'s up to the scale $\Lambda \approx 4\pi v \approx 3$ TeV and may give an important contribution to the ElectroWeak Precision Tests (EWPT), both at tree and loop level~\cite{HeavyV}.
For the composite fermions, which are the focus of this work, we consider Singlets, Doublets or Triplets under $SU(2)_{L+R}$. They also contribute, at loop level, to the EWPT. Furthermore they carry a generation index $i=1,2,3$ as the standard elementary fermions and a $X$ quantum number that allows them to mix with the elementary fermions consistently with gauge invariance. We hope that the multiplicity of these states be somehow explained by the strong dynamics. A crucial assumption we make is that, in absence of this mixing, the elementary fermions are massless\footnote{An early paper discussing the mass generation of the ordinary fermions by mixing with composite fermions is~\cite{Kaplan:1991dc}.}  and the strong dynamics does not distinguish the different generations. We shall show that  a definite symmetry pattern for the mixing between the elementary and the composite fermions  allows to keep, in the Singlet case, the CKM picture of flavour physics, although with  some characteristic effects in flavour physics still emerging.

While not resting on any explicit model of EWSB, which may actually be difficult to exhibit at all with present knowledge, the interesting aspect of the picture outlined above is that some of its consequences can be analyzed on general grounds. 
Not surprisingly, this is based  on the symmetries of the problem: on one side the {\it chiral} $SU(2)_L\times SU(2)_R$ spontaneously broken down to $SU(2)_{L+R}$ and, on the other side, the postulated flavour symmetry, $G_f$, of the strong dynamics and of the {\it weak} gauge interactions in absence of mixing between the composite and the elementary fermions.

The content of the paper is the following. In Section 2 we describe the effective chiral Lagrangians of the composite fermions up to one derivative terms, including the mass mixing with the elementary fermions. In Section 3 we discuss some of the effects of the composite Doublets or Triplets in the EWPT. We take these considerations as evidence for a generic difficulty of Doublets or Triplets in a truly strongly interacting theory of EWSB, unless some parameters are suitably adjusted.  As shown in Section   4 , even considerations of flavour physics make Doublets and Triplets somewhat disfavoured. On the other hand, the effective Lagrangian for composite Singlets up to one derivative terms is identical to the one of elementary fermions with the same weak quantum numbers, thus screening at all their composite nature. They may play however an interesting role in Flavour Physics by mixing with the elementary fermions in a suitable way, as discussed in Section 4. 
We require that the flavour symmetry $G_f^{\mathcal{S}}$ be broken by mixing parameters which, treated as spurions, have definite transformation properties under $G_f^{\mathcal{S}}$. In the case of composite Singlets, this enables us to enforce a specific case of Minimal Flavour Violation, with the CKM matrix as the only control of flavour changing phenomena. Yet the mixing between the elementary and the composite singlets gives rise to significant residual effects both in the EWPT and in Flavour Physics, as discussed in Section 5.  While all the discussion is concentrated on quarks, the picture is easily extended to leptons (See Section 6). The collider phenomenology is briefly described in Section 7. The conclusions are summarized in Section 8.

%






%

\section{Effective Lagrangians}%

We consider in turn the effective Lagrangians for Singlets, Doublets and Triplets under $SU(2)_{L+R}$ up to one derivative terms, including the general mass mixings with the elementary fermions. We concentrate on quarks, leaving the straightforward extension to leptons in Section 7. As mentioned, the symmetry includes a $U(1)_X$ to allow the standard gauging of hypercharge.

Other than (\ref{U}), the key ingredient to describe the transformation properties of the various fields under the full $SU(2)_L\times SU(2)_R$ is  the \textit{little} matrix $u$ \cite{coleman} via%
\begin{equation}
U=u^{2}~.\nonumber
\end{equation}
This matrix parametrizes the $SU(2)_{L}\times SU(2)_{R}/SU(2)_{L+R}$ coset and
transforms as\footnote{Note that here as in~\cite{HeavyV} we follow the convention, usually adopted in QCD, where the $L$-transformations act on the right and viceversa. This is unlike what is normally done in discussing strong EWSB.}
\begin{equation}
u\rightarrow g_{R}uh^{\dagger}=hug_{L}^{\dagger}~,\nonumber
\end{equation}
where $h=h(u,g_{L},g_{R})$ is uniquely determined by this equation.  In turn, important functions of $u$ are
\begin{equation}
\Gamma_{\mu}=\frac{1}%
{2}\left[  u^{\dagger}(\partial_{\mu}-i\hat{B}_{\mu})u+u(\partial_{\mu}%
-i\hat{W}_{\mu})u^{\dagger}\right]  ,\quad\Gamma_{\mu}^{\dagger}=-\Gamma_{\mu
}~,\quad\Gamma_{\mu}\rightarrow h \Gamma_{\mu} h^\dagger + h\partial_{\mu}h^\dagger~,
\end{equation}
and
\begin{equation}
u_{\mu}=iu^{\dagger}D_{\mu}Uu^{\dagger}=u_{\mu}^{\dagger},~\qquad u_{\mu
}\rightarrow hu_{\mu}h^{\dagger}~,
\end{equation}
where
\be
D_{\mu}U=\partial_{\mu}U-i\hat{B}_{\mu}U+iU\hat{W}_{\mu}~,\qquad\hat
{W}_{\mu}=gT^{a}W_{\mu}^{a}~,\qquad\hat{B}_{\mu}=g^{\prime}T^{3}B_{\mu}~.
\label{def}%
\ee

\subsection{Singlets}

Singlets under $SU(2)_{L+R}$ that can mix with the standard quarks are $U$ and $D$, carrying colour and $X$-numbers $2/3$ and $-1/3$ respectively. Their covariant kinetic term is the trivial one and no other one-derivative term appears (Tr($u_\mu$)=0). 
In absence of mixing with the elementary fermions, the interactions of the singlets $\mathcal{S}$ with the Goldstone fields appear only at two-derivative level and are suppressed by inverse powers of the cutoff.
Their composite nature may be difficult to see directly. They may however play an interesting role in giving masses by mixing to the standard fermions. Introducing the fictitious doublets, one per generation,
\be
Q    \equiv\left(
\begin{array}
[c]{c}%
U\\
D
\end{array}
\right) 
\ee
the most general mixing mass term, including the necessary $SU(2)_R$ 
breaking, is\footnote{The $U$ appearing in all the $L_{mix}$ of this Section is the matrix in (\ref{U}) and should not be confused with the heavy fermion $U$.}
%
%
%
%
%
\be
L_{mix}^\mathcal{S}  = m_L^u \bar{Q}_R  \Pu U q_{L}+  m_R^u \bar{Q}_L  \Pu q_{R} 
+ m_L^d\bar{Q}_R  \Pd U q_{L}+  m_R^d\bar{Q}_L  \Pd q_{R} + {\rm h.c. }
\label{LmixS}
\ee
where $\Pud = (1\pm \sigma_3)/2$.
All these $m$'s are matrices in flavour space. 
In absence of mixing with the elementary fermions, 
the composite singlets have flavour independent masses:
\be
L_{\mathcal{S}} =  i\bar{Q}\gamma^{\mu}(\partial_{\mu}-i g^\prime X B_\mu) Q
+ M_U\bar{U} U + M_D \bar{D} D~.
\label{eq:heavymasses}
\ee
A discussion about the flavour structure of the model 
is postponed to Section~\ref{sect:Flav} 
and~\ref{sect:Intra}. Here we limit to note that
in the mass-eigenstate basis the relation between the mass 
parameters in (\ref{LmixS})--(\ref{eq:heavymasses}) 
and the physical masses of light and heavy states are given 
by Eqs.~(\ref{eq:newl1})--(\ref{eq:newl3}).


\subsection{Doublets}

The Doublets 
\be
\mathcal{D}    \equiv\left(
\begin{array}
[c]{c}%
T\\
B
\end{array}
\right),
\ee
of $X$-number $1/6$, transform under $SU(2)_{L+R}$ as $\mathcal{D}\rightarrow h\mathcal{D}$. We call the {\it up} and {\it down} components of $\mathcal{D}$ respectively $T$ and $B$, not to confuse them with the $SU(2)_{L+R}$ singlets with the same charge, $U$ and $D$.
The most general invariant Lagrangian up to one-derivative terms is
\begin{equation}
L_{\mathcal{D}} = i\bar{\mathcal{D}}\gamma^{\mu}(\partial_{\mu}+\Gamma_{\mu}-i g^\prime X B_\mu)\mathcal{D}+\frac{\alpha}{2}\bar{\mathcal{D}}\gamma_{\mu}\gamma_{5}u_{\mu}\mathcal{D} + M_{\mathcal{D}}\bar{\mathcal{D}}\mathcal{D}.
\end{equation}
The strong dynamics is assumed to conserve parity.

The mixing Lagrangian is
\be
L_{mix}^\mathcal{D}  = m_L^u \bar{\mathcal{D}}_R 
u^\dagger \Pu U q_{L}+  m_R^u \bar{\mathcal{D}}_L u^\dagger \Pu q_{R} +
 m_L^d\bar{\mathcal{D}}_R u^\dagger \Pd U q_{L}+  m_R^d \bar{\mathcal{D}}_L u^\dagger \Pd q_{R} + {\rm h.c.} \\
\ee
whose invariance follows from 
$u^{\dagger}\mathcal{D}    \rightarrow g_L u^{\dagger}\mathcal{D}$ and 
$u \mathcal{D}     \rightarrow g_R u \mathcal{D}$.
Neglecting terms containing pions, 
\be
i \Gamma_{\mu} \rightarrow \frac{1}{2} (\hat{W}_\mu +  \hat{B}_\mu)~,\quad
u_\mu \rightarrow   \hat{B}_\mu - \hat{W}_\mu~.
\ee
Therefore the electroweak interactions of the Doublets are
\begin{equation}
L_\mathcal{D}^{int} = \bar{\mathcal{D}}\gamma^{\mu} [\hat{W}_\mu \frac{1}{2} (1-\alpha \gamma_5) + \hat{B}_\mu \frac{1}{2} (1+\alpha \gamma_5) + g^\prime X {B}_\mu] \mathcal{D}
\label{DEW}
\end{equation}
(which reduce to the ones of an elementary quark family for $\alpha=1$).

\subsection{Triplets}

The Triplets $\mathcal{T}$ that contain bot an {\it up} and a {\it down}-type quark have $X=2/3$ or $X=-1/3$ and transform under 
$SU(2)_{L+R}$ as $\mathcal{T}\rightarrow h\mathcal{T} h^+$.
In  $2\times 2$ matrix notation, e.g. for  $X=2/3$, it is
\[
\mathcal{T}=\left(
\begin{array}
[c]{cc}%
T/\sqrt{2} & X^{5/3}\\
B & -T/\sqrt{2}
\end{array}
\right)
\]
where $X^{5/3}$ is an exotic quark of charge $2/3$.

The most general invariant  Lagrangian is
\be
L_{\mathcal{T}}    =i Tr[ \bar{\mathcal{T}}\gamma^{\mu}(\partial_{\mu}\mathcal{T} +[\Gamma_{\mu}, \mathcal{T}] -i g^\prime X {B}_\mu \mathcal{T})]
+\alpha Tr[ \bar{\mathcal{T}}\gamma_{\mu}\gamma_{5}u_{\mu}\mathcal{T}] + M_\mathcal{T} \bar{\mathcal{T}}\mathcal{T},
\end{equation}
with the electroweak gauge interactions that reduce to
\begin{equation}
L_{\mathcal{T}}^{int} = Tr[ \bar{\mathcal{T}} \gamma^{\mu} (\hat{W}_\mu (1 -\alpha \gamma_5)  + \hat{B}_\mu (1 +\alpha \gamma_5)+ g^\prime X {B}_\mu) \mathcal{T}].
\label{TEW}
\end{equation}

To construct the mixing Lagrangian, note that, under $SU(2)_L\times SU(2)_R$
\begin{align*}
u^{\dagger}\mathcal{T}u^{\dagger}  &  \rightarrow g_L(u^{\dagger}\mathcal{T}u^{\dagger})g_R^{\dagger
}\\
u\mathcal{T}u^{\dagger}  &  \rightarrow g_RR(u\mathcal{T}u^{\dagger})g_R^{\dagger}.%
\end{align*}
Defining the vectors
\begin{equation}
(u\bar{\mathcal{T}} u^{\dagger})_{1i} = \bar{v}_{i},~~
(u\bar{\mathcal{T}}u)_{1i} = \bar{w}_{i}
\end{equation}
the mixing Lagrangian is


\be
L_{mix}^{\mathcal{T}}   = m_L^u\bar{w}_R U^\dagger \Pu U q_{L}
+ m_R^u \bar{v}_L  \Pu q_{R} +
 m_L^d \bar{w}_R   U^\dagger \Pd U q_{L}
+ m_R^d \bar{v}_L  \Pd q_{R} + {\rm h.c.}
\ee


\section{EWPT for Doublets and Triplets}



\subsection{$\Delta S$}
\label{sect:DS}

Even in absence of any mixing, the Doublets and the Triplets contribute to the the EWPT through the $S$-parameter.
Each doublet and  triplet contribute respectively to the $S$-parameter as
\be
\Delta S^{({\mathcal D})} = \frac{1}{2\pi}[1-(\alpha^2-1)( \log{\Lambda^2/M_{\mathcal{D}}^2})]
\label{S(doublet)}
\ee
\be
\Delta S^{({\mathcal T})} = \frac{2}{\pi}[1-(\alpha^2-1)( \log{\Lambda^2/M_\mathcal{T}^2})]
\label{S(triplet)}
\ee
where $\Lambda$ is a suitable UV cutoff.
Especially if there is one such contribution per generation, this is a pretty large effect (which  might be negative if $\alpha >1$). This $\Delta S$ is reminiscent of  the well known contribution from technifermions in TechniColour. In our "effective" view,  there is one  contribution as in (\ref{S(doublet)}) or (\ref{S(triplet)}) per generation and we would have to add it to the contribution from heavy vectors, which occurs at tree level.

Eq.s (\ref{S(doublet)}) or (\ref{S(triplet)}) appear to us as a generic difficulty for a truly strongly interacting theory of EWSB. It is of some interest, however, to make contact with  models in the literature that have a moderate or even vanishing $\Delta S$ from fermion loops. One case is if the 
 composites, either doublets or triplets, occur in full representations of $SU(2)_L\times SU(2)_R$ before symmetry breaking. This is meaningful only 
if the compositeness scale is 
higher than the scale of electroweak symmetry breaking,
like in models where there is a Higgs doublet, either elementary or composite. Examples are vector-like representations $(2,1) \oplus (1,2)$ or $(2,2)$ of
$SU(2)_L\times SU(2)_R$ with suitable $X$-numbers, 
 giving respectively 2 doublets $(\mathcal{D}_1, \mathcal{D}_2)$ or   one triplet and one singlet ($\mathcal{T} \oplus S$) under $SU(2)_{L+R}$.
 In this case, $\Delta S$ may vanish. As easily seen by explicit calculations, this requires degeneracy of the full representation after $SU(2)_L\times SU(2)_R\rightarrow SU(2)_{L+R}$ breaking and perturbative uncorrected gauge couplings, which speaks against a strongly interacting composite Higgs boson.

Another  example of formally vanishing $\Delta S$ from fermion loops comes from a Dirac fermion transforming as  $(1,2,1)$ under $SU(2)_L\times SU(2)_C \times SU(2)_R$, broken down to the diagonal subgroup, $SU(2)_{L+C+R}$, like in the so called {\it Three-site Model}~\cite{Threesite}. 
This requires that the mass of the doublet be much larger than the masses of the vector bosons and that perturbation theory makes sense in all the couplings, including the coupling $g_C$  of $SU(2)_C$. In this case, (\ref{S(doublet)}) remains formally correct, with $\alpha=0$, but the full $\Delta S$ from the fermionic loop is reabsorbed, after renormalization of $g_C$,  in the tree level effect to $S$ due to the kinetic mixing of the vector bosons.

\subsection{$\Delta T$ and $Z\rightarrow \bar{b} b$}
\label{eq:Zbb}

Unlike $\Delta S$, a contribution to the $T$ parameter arises only after breaking of the custodial symmetry from the composite/elementary mixing, which is important from the third family only. In fact, in presence of a strong breaking of this symmetry in the left-handed sector, i.e. 
 $m^d_L << m^u_L$, explicit calculations show that  $\Delta T$ is always unacceptably large, whereas it is moderate and generally positive, 
if $m^u_L\approx m^d_L$, so as to minimize custodial breaking. In this case, however, it is crucial to watch the $Z\bar{b}_L b_L$ coupling,  because of the  $b_L/B_L$-mixing that occurs at tree level.

Let us consider the deviations from the SM of the $Z\bar{b}b$ couplings, $\delta g_L, \delta g_R$.
It is in general (an overall factor $g/c_W$ taken away)
\be
\delta g_{L,R} = (s^b_{L,R})^2 [g_{L,R} (B) - g_{L,R} (b)]
\ee
where $s^b_{L,R}$ are the sines of the mixing angles in the down sector
for the third generation. 
From the Lagrangians of Eq.s (\ref{DEW}) and (\ref{TEW}) one gets respectively:

\begin{itemize}
\item Doublet case
\be
\delta g_{L} = \frac{(s^b_L)^2}{4}  (1 - \alpha)~,  \qquad
\delta g_{R} = -\frac{(s^b_{R})^2}{4} (1 - \alpha)~.
\ee
\item Triplet case 
\be
\delta g_{L} = -\frac{(s^b_L)^2} {2} \alpha~, \qquad 
\delta g_{R} = -\frac{(s^b_{R})^2} {2} (1-\alpha)~. 
\ee
\end{itemize}
Now for generic 
$\alpha = O(1)$ and $m^d_L \approx m^u_L$ for the third generation, this is a strong constraint. 
In fact, from the diagonalization of the mass matrix of the third generation (see Section~5) one finds 
\be
 s^b_L \approx \frac{m_t}{M_T} \frac{1}{s_R^t c_R^t }~,  \qquad 
 s^b_R \approx \frac{m_b}{m_t} s_R^t~,
\ee
where $s_R^t$ and $c_R^t$ denote sine and cosine of the 
mixing angle of the right-handed top with its 
composite partner of mass $M_T$. Whatever the value of $s^t_R$ is, 
these equations make $\delta g_{R} $ irrelevant (i.e.~no explanation  
offered for the notorious problem of the $b$ forward-backward asymmetry) 
whereas, depending on the value of $\alpha$, a strong bound on $M_T$ generally 
arises from the contribution of $\delta g_{L}$ to $\Gamma (Z\rightarrow \bar{b}b)$\footnote{
$\delta g_L =0$ or $\alpha=0$ in the Triplet case results if the Triplet comes from a $(2,2)$ of $SU(2)_L\times SU(2)_R$ and the perturbative gauge couplings are kept uncorrected~\cite{Agashe:2006at}.}.

From the composite/elementary mixing also the 
Singlets contribute  to $\Delta T$ 
and to $Z\rightarrow \bar{b} b$. We shall come back to this in Section 5.

\section{Flavour symmetries}
\label{sect:Flav}

As mentioned in the Introduction we assume that, in absence of composite/elementary mixing, the system possesses a large flavour symmetry which extends the one of the SM for vanishing Yukawa couplings, 
\be
G_f^{SM} = SU(3)_q\times SU(3)_{uR}\times SU(3)_{dR},
\ee
to include also the flavour symmetry of the composite sector, i.e.
\be
G_f^{\mathcal{S}} = SU(3)_U\times SU(3)_D\times SU(3)_q\times SU(3)_{uR}\times SU(3)_{dR}
\ee
or
\be
G_f^{\mathcal{D}, \mathcal{T}} = SU(3)_{\mathcal{D}, \mathcal{T}}\times SU(3)_q\times SU(3)_{uR}\times SU(3)_{dR}.
\label{GfC}
\ee
Both in the SM viewed as an effective theory or here,  these flavour symmetries have to be broken appropriately to keep consistency with experiments. An additional problem of the flavour-breaking mixing terms in the effective Lagrangians in Section 2, if treated generically, is the large number of physical parameters, even increased relative to the SM.

\subsection{Minimal Flavour Violation}

As  well known, the SM model viewed as an effective theory
gives a consistent description of flavour physics,  provided the flavour group in the quark sector, $G_f^{SM}$, is only broken by two dimensionless parameters, $\YU$ and $\YD$, which, treated as spurions, transform as
\be
\YU = (3, \bar{3}) \quad {\rm under} \quad SU(3)_{uR} \times SU(3)_{q}
\ee
\be
\YD = (3, \bar{3}) \quad {\rm under} \quad SU(3)_{dR} \times SU(3)_{q}
\ee
Such an hypothesis enforces in particular the successful CKM picture, since, without loss of generality,  $\YD$ can be reduced to diagonal form, 
$\YD = \lambda^d$, and similarly   $\YU$ can be diagonalized 
up to a single unitary matrix, $\YU = \lambda^u V$, 
where $V$ is the CKM matrix. As long as this symmetry and this symmetry breaking pattern is respected, even the inclusion of higher dimensional operators, 
suppressed by a scale of $3\div 5$ TeV, is harmless~\cite{MFV}.

\subsection {Singlets}
\label{sect:Sflavour}

It would be nice if the above picture of flavour physics could be extended to the situation we are considering here, keeping in particular under control the number of new parameters in the flavour sector. This is in fact neatly the case under one of the 
 following circumstances for the flavour group $G_f^{\mathcal{S}}$, that we name {\it Parity Conserving (PC)} and {\it Parity Breaking (PB)}:
\begin{itemize}
\item 1. {\it Parity Conserving}
\end{itemize}
$G_f^{\mathcal{S}}$ is only broken by
\be
\YU_1 = (3, \bar{3}) \quad {\rm under} \quad SU(3)_U \times SU(3)_{q+uR}
\ee
\be
\YD_1 = (3, \bar{3}) \quad {\rm under} \quad SU(3)_D \times SU(3)_{q+dR}
\ee
where $SU(3)_{q+uR}$ or $SU(3)_{q+dR}$ denote the corresponding diagonal groups. For the mass matrices in Eq.~(\ref{LmixS}) this implies:
\be
m_L^u = v \YU_1,~~m_R^u = f^u \YU_1,~~m_L^d = v \YD_1,~~m_R^d = f^d \YD_1,
\ee
where $f^u, f^d$ are two mass scales, likely of the same order as the composite quark masses, $M_U, M_D$ in (\ref{eq:heavymasses}), 
and $V$ is again the CKM matrix.
After diagonalization of $\YU_1$ and $\YD_1$ and suitable redefinitions of the various fields in generation space, the overall mass Lagrangian can be written as 
\be
L_{mix}^{\mathcal{S}}(PC)  = v  \Bar{U}_R \lambda^u V u_L + f^u \Bar{U}_L \lambda^u u_R +
  v \Bar{D}_R \lambda^d d_L + f^d \Bar{D}_L \lambda^d d_R + {\rm h.c.}
\ee
with diagonal $\lambda^u$ and $\lambda^d$.
\begin{itemize}
\item 2. {\it Parity Breaking }
\end{itemize}
$G_f^{\mathcal{S}}$ is broken down to $SU(3)_{U+uR}\times SU(3)_{D+dR} \times SU(3)_q$, which is in turn only broken by
\be
\YU_2 = (3, \bar{3}) \quad {\rm under} \quad SU(3)_{U+uR}\times SU(3)_{q}
\ee
\be
\YD_2 = (3, \bar{3}) \quad {\rm under} \quad  SU(3)_{D+dR}\times SU(3)_{q},
\ee
so that in this case
\be
m_L^u = v \YU_2,~~m_R^u = f^u {\bold 1},~~m_L^d = v \YD_2,~~m_R^d = f^d {\bold 1}.
\ee
The mass Lagrangian can be written as
\be
L_{mix}^{\mathcal{S}}(PB)  = v  \Bar{U}_R \lambda^u V u_L + f^u \Bar{U}_L u_R +
  v \Bar{D}_R \lambda^d d_L + f^d \Bar{D}_L d_R + {\rm h.c.}  
\ee

Note that in both cases the direct mass terms between two elementary fermions is not compatible with the required symmetry and symmetry breaking.

A third case analogous to 2 above with $G_f^\mathcal{S}$  broken down to $SU(3)_{U+q}\times SU(3)_{D+q} \times SU(3)_{uR}\times SU(3)_{dR}$ would imply
\be
m_L^u \propto {\bold 1},~~m_R^u = f^u Y^u,~~m_L^d \propto {\bold 1},~~m_R^d = f^d Y^d.
\ee
Although also leading to a case where flavour mixing 
is controlled only by the CKM matrix, this case
is not compatible with observations 
unless the mass $M_U$ of the composite $U$-quarks is taken well above 5 TeV.  This is due to the intra-generation mixing of the left-handed light
quarks, which spoils the precise tests of 
CKM unitarity~(see e.g.~Ref.~\cite{Antonelli:2008jg}).

\subsection {Doublets and Triplets}
\label{sect:DTflav}
It is of interest to ask if suitable symmetry conditions, analogous to the previous ones,  can force the CKM picture of flavour physics also in the case of composite doublets or triplets. The answer is no\footnote{As in the singlet case we discard here as well, and for the same reason, the symmetry $SU(3)_{\mathcal{D}+q} \times SU(3)_{uR}\times SU(3)_{dR}$.}.
The overall flavour symmetry now is $G_f^{\mathcal{D}, \mathcal{T}}$ in (\ref{GfC}). Therefore, with reference, e.g., to the PC case above, the analogous condition is that $G_f^{\mathcal{D}, \mathcal{T}}$ be only broken by
\be
\YU_3 = (3, \bar{3}) \quad {\rm under} \quad
SU(3)_{\mathcal{D}, \mathcal{T}} \times SU(3)_{q+uR}
\ee
\be
\YD_3 = (3, \bar{3}) \quad {\rm under} \quad
SU(3)_{\mathcal{D}, \mathcal{T}} \times SU(3)_{q+dR}
\ee
In turn, by suitable redefinitions of the various fields, the mixing Lagrangian can be written as
\be
L_{mix}^{\mathcal{D}, \mathcal{T}}(PC) = v  \Bar{U}_R \mathcal{V} \lambda^u V u_L + f^u \Bar{U}_L \lambda^u u_R +
  v \Bar{D}_R \lambda^d d_L + f^d \Bar{D}_L \lambda^d d_R + {\rm h.c.}
\ee
where $\mathcal{V}$ is a further unitary matrix that cannot be rotated away from the overall Lagrangian. 

\subsection{Flavour breaking by higher dimensional operators}
\label{sect:HDops} 

As in the case without composite fermions, we have to ask if higher dimensional operators consistent with the symmetries and the symmetry breaking described above are compatible with observations. There are two sets of flavour-changing neutral-current (FCNC)
dimension-six operators, weighted by the inverse square of the cutoff and by  
suitable dimensionless coefficients, $c_{ij}$, $i,j=1,2,3$:
\be
O_{LL}^{ij} =  (\bar{q}_L^i \gamma_\mu q_L^j)^2,~~~
O_{LR}^{ij} = (\bar{q}_R^i \Pd U [\gamma_\mu,\gamma_\nu] q_L^j) B_{\mu \nu}
\ee
which are most significant. In Table~\ref{tab:FCNCs} we give the expected expressions for the $c_{ij}$ in the Singlet case according to the symmetry breaking 
conditions formulated in Section~\ref{sect:Sflavour}
and we compare them with the analogous expressions 
obtained in MFV~\cite{MFV}.
They are numerically equivalent  if $s_R^t$ is of order unity, 
showing that a cutoff scale of about $3\div 5$ TeV 
is also in this case compatible with current data.

The situation is different in the Doublet or Triplet cases. 
Following the discussion in Section~\ref{sect:DTflav}, 
a most dangerous effective Lagrangian is
\be
\Delta L^{\Delta F=2}_{\mathcal{D}, \mathcal{T}} = 
\frac{f_d}{\Lambda^3} (\bar{q}_R \Pd \YDdag_3 
\YU_3 U q_L)^2
\quad 
\rightarrow 
\quad 
\frac{m_d m_c  (\mathcal{V}_{12})^2 }{v^2}
\frac{M_D M_U}{f^u \Lambda^3} (\bar{d}_R s_L)^2
\ee
where in the last step we have selected the $\Delta S=2$ contribution 
and $m_d, m_c$ are the masses of the down and charmed quarks. The matrix element 
of this effective Lagrangian
between neutral kaons,  for $ |\mathcal{V}_{12}^2| \approx 1$,  
is about 100 times bigger than the matrix element of the leading 
dimension-six $\Delta S=2$ operator in MFV.
This issue (and a similar difficulty in $\Delta S=1$ left-right 
operators) is a manifestation of a general problem
of composite models~\cite{Davidson:2007si}:
it appears in all cases where the suppression of FCNCs
is attributed only to the small mixing of SM fermions 
and heavy states, but there is no flavour 
alignment between light and heavy states 
(see e.g.~\cite{Contino:2006nn}).

\begin{table}[t]
\begin{center}
\begin{tabular}{|c||c|c|c|} 
\hline
Operators  &  MFV &  Singlets PC & Singlets PB \\ \hline\hline
\raisebox{0pt}[20pt][10pt]{$ \dis\frac{c_{ij}}{\Lambda^2} O_{LL}^{ij}$ } 
 &  $c_{ij}=(\YUdag \YU)_{ij}^2$  &  $c_{ij}=(\YUdag_1 \YU_1)_{ij}^2 $  &   
    $c_{ij}=(\YUdag_2 \YU_2)_{ij}^2 $  \\
\raisebox{0pt}[0pt][15pt]{ $(\Delta F=2)$ } 
 &  $ \approx  \dis\frac{m_t^4}{v^4} (V^*_{3i} V_{3j})^2 $  &  
    $ \approx  \dis\frac{m_t^2}{v^2} \frac{M_T^2}{(f^u)^2} (V^*_{3i} V_{3j})^2 $  &  
    $\approx  \dis\frac{m_t^4}{v^4}  \frac{M_T^4}{(f^u)^4} (V^*_{3i} V_{3j})^2 $  \\
\hline 
\raisebox{0pt}[20pt][10pt]{$\dis\frac{c_{ij}}{\Lambda^2} O_{LR}^{ij}$ }  
 &  $c_{ij}=(\YD\YUdag \YU)_{ij} $ &  $c_{ij}=\frac{f^d}{\Lambda} (\YDdag_1 \YD_1 \YUdag_1 \YU_1)_{ij} $
 &  $c_{ij}=\frac{f^d}{\Lambda} (\YD_2 \YUdag_2 \YU_2)_{ij} $ \\
\raisebox{0pt}[0pt][15pt]{ $(\Delta F=1)$ }
 &  $ \approx\dis\frac{m_{d_i} m_t^2}{v^3}  V^*_{3i} V_{3j} $ &
    $ \approx \dis\frac{m_{d_i} m_t}{v^2}  \frac{M_T M_D}{ f^u \Lambda} 
     V^*_{3i} V_{3j}$  &  
    $ \approx \dis\frac{m_{d_i} m_t^2}{v^3 }  \frac{M_T^2 M_D }{(f^u)^2\Lambda} 
     V^*_{3i} V_{3j}$  \\
\hline
\end{tabular}
\end{center}
\caption{\label{tab:FCNCs} Comparison of the coefficients of 
the leading dimension-six operators relevant to 
$\Delta F=2$ and $\Delta F=1$ FCNC transitions of down-type quarks.}
\end{table}

\section{Composite/elementary mixing effects for Singlets}
\label{sect:Intra}

\subsection{Tree level}

After sending $u_L \rightarrow V^\dagger u_L$, both the up and down mass matrices 
reduce to three $2\times 2$ blocks, each labelled by a generation index $i$: 
\be
M^{(u)}_i = \left(
\begin{array}
[c]{cc}%
0 & f^u \lambda^u_i~ [ f^u ] \\
 v \lambda^u_i~   & m_U
\end{array}
\right)~, 
\qquad
M^{(d)}_i =\left(
\begin{array}
[c]{cc}%
0 & f^d \lambda^d_i~ [ f^d ] \\
v \lambda^d_i~  & m_D
\end{array}
\right)~,
\ee
where the values outside/within square brackets correspond to the 
PC/PB cases of Section~\ref{sect:Sflavour}.
Relative to the SM, this introduces two extra parameters for each 
quark, which can be chosen as the mass of the heavy partner 
and the mixing angle in the left-handed or in the 
right-handed sector. In the limit where 
we neglect light quark masses, the corresponding  left-handed
mixing angles can be set to zero and the heavy states 
decouples in low-energy observables. 

Considering a generic $2\times2$ block, and defining left and 
right mixing angles as follows, 
\be
\left(\ba{cc}
 -c^q_R &  s^q_R \\
  s^q_R &  c^q_R \ea\right) 
\left(\ba{cc}
0 & m^q_R\\
 m^q_{L} & M_Q \ea\right)
\left(\ba{cc}
 c_L^q & s_L^q \\
-s_L^q & c_L^q \ea\right)~=~{\rm diag}(m_q, M_{Q_q})~, 
\label{eq:newl1}
\ee
we have ($q$ stands for $u$ and $d$ and we omit for simplicity the index $q$ on the r.h.s.): 
\bea
t^q_{R(L)} &=& \frac{s^q_{R(L)}}{c^q_{R(L)}} ~=~  \frac{m_{R(L)}^2-m_{L(R)}^2
 -M_Q^2+\sqrt{M_Q^4+2M_Q^2(m_L^2+m_R^2)+(m_R^2-m_L^2)^2} }{2 m_{R(L)} M_Q}~,     \no\\
M^2_{Q_q}(m_q^2) &=& \frac{M_Q^2+m_{R}^2+m_{L}^2 \pm
 \sqrt{M_Q^4+2M_Q^2(m_L^2+m_R^2)+(m_R^2-m_L^2)^2} }{2}~, 
\label{eq:newl2}
\eea
which in the limit $m^q_L \ll M_Q$ reduce to
\be
M_{Q_q} \approx \sqrt{M_Q^2+(m^q_R)^2}~,\qquad 
m_q \approx \frac{m^q_L m^q_R}{M_{Q_q}}~,\qquad 
s^q_R \approx \frac{m^q_R}{M_{Q_q}}~,\qquad 
s^q_L \approx \frac{m^q_L}{M_{Q_q}} c_R^q~.
\label{eq:newl3}
\ee
The results in Eq.~(\ref{eq:newl1})--(\ref{eq:newl3}) are completely 
general (they holds also for Doublets and Triplets in the appropriate 
mass-eigenstate basis). In the two specific cases discussed in
Section~\ref{sect:Sflavour} they imply
\be
(s^q_L)^2 |_{\rm PC} 
\approx 
\frac {v m_q M_Q^2 }{f^q M^3_{Q_q} } 
\qquad {\rm or}\qquad
(s^q_L)^2  |_{\rm PB}
\approx  \frac {m^2_q M_Q^2 }{(f^q)^2 M^2_{Q_q}} ~.
\label{seni}
\ee

Since the two right-handed fields have the same quantum numbers, 
the rotation in the right-handed sector does not lead to  
observable effects and we can eliminate it 
by means of the exact relation 
\be
t^q_R t^q_L = \frac{m_q}{M_{Q_q}}~.
\ee

Both in the PC and in the PB case the right-handed mixing in  the top
sector can be large if $f^u \sim M_U$. As we have seen in Section~\ref{sect:HDops} 
this configuration is required for a natural suppression of the 
dimensions-six FCNC effective operators. We shall similarly assume $f^d \sim M_D$.

The rotation to the mass eigenstates leads to 
modifications in the interaction part of the SM Lagrangian.
The couplings of light and heavy fermions 
to Goldstone bosons, $W$ and $Z$ fields 
(left-handed component) can be obtained from 
the SM Lagrangian with the replacements
\be 
q^i_L \to c^i_L q^i_L +s^i_L Q^i_L~, \qquad 
m_i q^i_R \to m_i c^i_L q^i_L + M_{Q_i} s^i_L Q^i_L~.
\ee
For instance, the currents coupled to $W$ and  $Z$ fields, 
written in terms of the mass eigenstates, are
\bea
J^\mu_W &=& \frac{g}{\sqrt{2}} \sum_{ij=1,3} V_{ij} 
\left[ c_L^{u_i}c_L^{d_j} \bar u^i_L \gamma^\mu  d^j_L
+ s_L^{u_i}s_L^{d_j} \bar U^i_L \gamma^\mu  D^j_L
+ c_L^{u_i}s_L^{d_j} \bar u^i_L \gamma^\mu  D^j_L
+ s_L^{u_i}c_L^{d_j} \bar U^i_L \gamma^\mu  d^j_L \right]~, \no \\
J^\mu_Z &=& \frac{g}{c_W} \sum_{i=1,6}
 \left\{ (T_3)_i \left[ (c^i_L)^2 \bar q^i_L\gamma^\mu q^i_L + 
 (s^i_L)^2 \bar Q^i_L\gamma^\mu Q^i_L +  s^i_L c^i_L \bar q^i_L\gamma^\mu Q^i_L 
+ s^i_L c^i_L \bar Q^i_L \gamma^\mu q^i_L \right] 
\right. \no \\ && \left. \qquad 
- Q_i s^2_W \left[ \bar q^i \gamma^\mu q^i  + \bar Q^i \gamma^\mu Q^i \right]
\right\}~,
\label{weakcurrents}
\eea
while the fermion couplings to a single
Goldstone boson field are
\bea
\delta L_{\pi} &=& 
\frac{ \sqrt{2}i \pi^+ }{v}  \sum_{ij=1,3} V_{ij}
\big[ c_L^{u_i} m_{u_i} \bar u_R^i  + s_L^{u_i} M_{U_i} \bar U_R^i \big]
\big[ c_L^{d_j} d_L^j  + s_L^{d_j}  D_L^j \big] \no\\
&+& 
\frac{ \sqrt{2} i \pi^- }{v}  \sum_{ij=1,3} V^*_{ji}
\big[ c_L^{d_i} m_{d_i} \bar d_R^i  + s_L^{d_i} M_{D_i}  \bar D_R^i \big]
\big[ c_L^{u_j} u_L^j  + s_L^{u_j}  U_L^j \big]~
\no  \\
&-& \frac{ i \pi^0 }{v}  \sum_{i=1,3} 
\big[ c_L^{d_i} m_{d_i} \bar d_R^i  + s_L^{d_i} M_{D_i}  \bar D_R^i \big]
\big[ c_L^{d_i} d_L^i  + s_L^{d_i}   D_L^i \big] \no\\
&+& 
\frac{ i \pi^0 }{v}  \sum_{i=1,3} 
\big[ c_L^{u_i} m_{u_i} \bar u_R^i  + s_L^{u_i} M_{U_i} \bar U_R^i \big]
\big[ c_L^{u_i} u_L^i  + s_L^{u_i} U_L^i \big]~
+~{\rm h.c.} 
\label{pi_int}
\qquad
\eea

For all the light quarks, including the $b$, the mixing 
angles are very small. In principle, one can expect 
some impact in the precise tests of CKM unitarity, 
which requires corrections to $V_{us}$ and $V_{ud}$ 
not exceeding $1\%$ and $0.1\%$, respectively. 
However, this condition turns out to be easily 
fulfilled even for small right-handed mixing.
Also the tree-level correction to 
the $b$-quark coupling  $g_L$,
discussed in Section~\ref{eq:Zbb},
turns out to be negligible for  
$f^d \sim M_D$.

\subsection{Loop effects}

The only significant impact of the light-heavy 
mixing at low energies arises at the
loop level from the top sector\footnote{For previous related calculations, see \cite{Carena:2007ua}, \cite{Barbieri:2007bh}, \cite{Lodone:2008yy}}. Here the sizable 
mixing with the heavy partner can lead to non-negligible
corrections  to the top-induced non-decoupling effects 
in FCNCs and EWPT. The corrections can be  
summarised as follows (see the Appendix for more details):
\begin{itemize}
\item{} SM amplitudes which have a finite limit 
for $x_t=m^2_t/m^2_W \to \infty$,
such as $\cA(b\to s\gamma)$, are modified in ($s=s_L^t$)
\be
F_{SM}(x_t) \to (1-s^2) F_{SM}(x_t) + s^2 F_{SM}(x_T)~,  \qquad x_T= M^2_T/m^2_W~.
\ee
\item{} SM amplitudes which grows linearly with $x_t$, such a 
$\delta T$, $\cA(Z\to \bar d^i d^j)$, $\cA(\Delta F=2)$,  receive a 
correction which is universal in the gauge-less limit ($m_W \to 0$):
\bea
 F_{\rm SM} (x_t) &\to&
 F_{\rm SM} (x_t) \times R(x_t,x_T) \no \\ 
R(x_t,x_T) &=& 1 + s^2 
\left[ -2  + s^2 \left(\frac{M_T^2}{m_t^2} +1\right) 
 +2 \frac{M_T^2}{M_T^2 -m_t^2}
\ln\left(\frac{m_T^2}{m_t^2}\right) 
 + \cO\left(\frac{m^2_W}{m_t^2},\frac{m^2_W}{M^2_T}\right)\right]~\ \qquad 
\label{eq:Runiv}
\eea
\end{itemize}

\begin{figure}[t]
\begin{center}
\includegraphics[width=9cm]{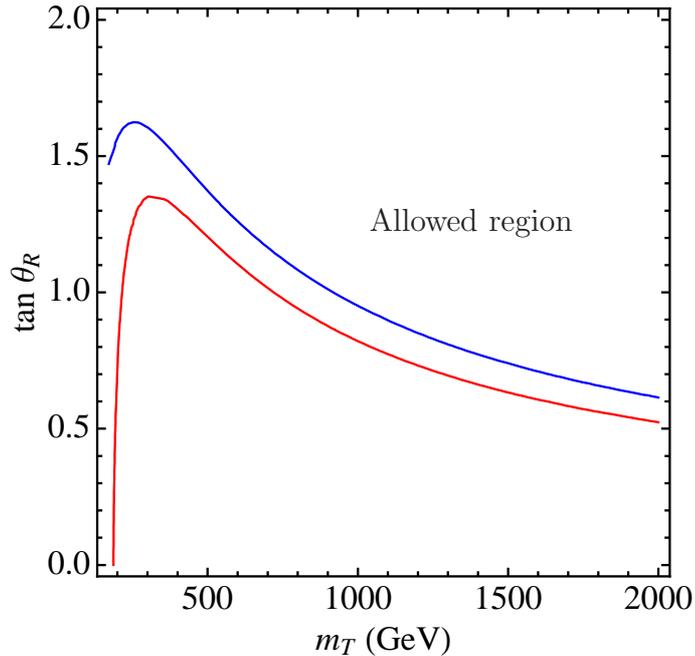}
\end{center}
\caption{Constraints in the $t_R$--$M_T$ plane at 95\% C.L.
from $\Delta F=2$ (red curve) and $Z \bar b b$ (blue curve).}
\label{fig:bounds}%
\vskip -8 cm 
\hskip 9 cm 
Allowed region
\vskip 8 cm
\end{figure}

The two most interesting phenomenological bounds, arising 
from $\Delta F=2$ (in particular $\epsilon_K$) and $Z\to b \bar b$,
are summarised in Fig.~\ref{fig:bounds}. 
In the case of $\Delta S=2$ we plot the constraint
following from $(\epsilon_K)_{\rm exp}/(\epsilon_K)_{\rm SM} =
0.92 \pm 0.14$, as derived in \cite{UTfit} from a recent
analysis of all $\Delta F=2$ amplitudes in the MFV 
framework. In the $Z\to b \bar b$ case we show the constraint
following from $(\delta g_L^b)_{\rm exp}/(\delta g_L^b)_{\rm SM} = 0.86 \pm 0.21$,
as derived by the $R_b$ measurement~\cite{EWWG},
where $\delta g_L^b$ is the deviation of $ g_L^b$
from its tree-level SM value.
The present constraints form $Z \to \bar b s$ and  $b\to s\gamma$
(see Ref.~\cite{Hurth:2008jc}) are substantially 
less severe.

As shown in Fig.~\ref{fig:bounds},
 $t^t_R = 1$
implies a lower bound around 800 GeV for $M_T$, but this limit evaporates as soon as $t^t_R\gtrsim 1.5$ .
Independently of the value of $M_T$ and $t^t_R$,  
the correlation of the various loop amplitudes, as described above, 
implies a small correction to $\delta T$: 
a positive contribution which does not exceed $\approx 10\%$
of $\delta T^{\rm SM}_{\rm top}$. This makes it unlikely that a too large contribution to the $S$-parameter can be reconciled with the EWPT by a significant $\delta T$ from composite singlets. Because of the 
different behaviour in the $m_t\to \infty$ limit,
the impact is much smaller in $b\to s\gamma$, 
where the positive correction does not exceed $\approx 2\%$.

\section{Leptons}

The picture described so far for the quarks
can be trivially extended to leptons with composites $E_i$ and  $N_i$, one per generation,  taking among the elementary leptons also the right-handed neutrinos $\nu_{R_i}$.  Note that the $N_i$ have no interaction at "renormalizable" level.
The smallness of the observed neutrino masses can be attributed to a large Majorana mass for the elementary $\nu_{R_i}$'s, $M$, related to the breaking of lepton number and much larger than the compositeness scale of the $N_i$, of mass $M_N$. 
This mass among elementary fermions can be present consistently with all the symmetries we have been talking about. Using a notation similar to the one of the quarks,  $m_L$ has to be sufficiently smaller than $M_N$, so that
 the light neutrinos are approximately $\nu_L + (m_L/M_N) N_L$ and have mass 
 \be
 m_\nu^{light} \approx (\frac{m_L}{M_N})^2 \frac{m_R^2}{M}.
 \ee

It may be interesting to study the phenomenology and the cosmology\footnote{As an example, the composite $N$ partner of a massless, or quasi massless neutrino, might be a Dark Matter candidate for suitable values of  its non renormalizable interactions.} of the  neutrino sector so extended, which is outside the scope of this work.

\section{Collider Phenomenology}

\begin{figure}[t]
\begin{center}
\includegraphics[width=9cm]{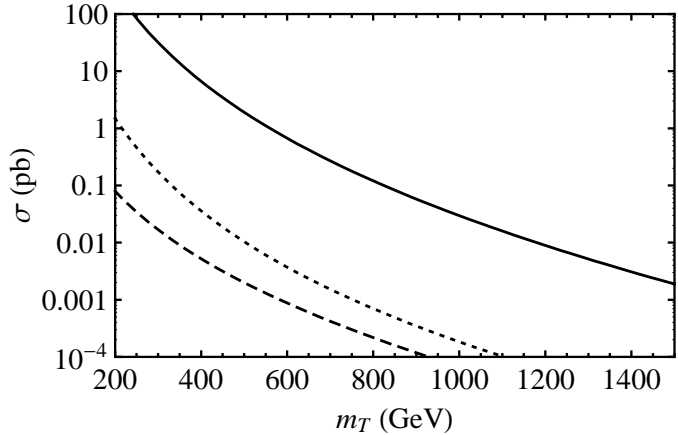}
\end{center}
\caption{ Production cross sections at LHC: for a pair of composite quarks of mass $M$ (full curve); for a singlet $T\bar{t}$ (dashed curve); for a singlet $T\bar{b}$ (dotted curve).
\label{fig:Xsections} }
\end{figure}

At the LHC, the pair production of any composite quark proceeds by gluon-gluon fusion. Single production, on the contrary, in association with an elementary quark is a weak Drell Yan process. 
In Fig.~\ref{fig:Xsections} we give both the pair-production cross section for a generic composite quark and the  $T\bar{t}$ and $T\bar{b}$ cross sections in the case of composite singlets, using the currents in (\ref{weakcurrents}). The single production cross sections are reduced by the small mixing angles (\ref{seni}), making all of them negligibly small  but the one of the $T$.

 The three singlets, of charge $2/3$ or $-1/3$, have splittings
 \be
 \frac{M_i-M_j}{\langle M \rangle}                    
\sim \left\{ \frac{m_i - m_j}{v},
\frac{(m_i^2-m_j^2)}{\langle M^2 \rangle}\right\}, 
 \ee
respectively in the PC and PB case,
where $m_i$ are the masses of the corresponding elementary quarks. 
Except the partner of the top in the PC case, all the other composites of given charge are highly degenerate.

The composite singlets have a narrow width, which is most easily computed by means of (\ref{pi_int}). The dominant decay of $U_i$ or $D_i$ is in the corresponding light state, $u_i$ or $d_i$ plus a $W$ or a $Z$, with
\be
\Gamma(U_i \rightarrow d_i+W) \approx  2 \Gamma(U_i \rightarrow u_i+Z) 
\approx \frac{1}{16\pi} ( s_L^{u_i} )^2 \frac{M_{U_i}^3 }{v^2}
\label{eq:BRZ1}
\ee
and, analogously,
\be
\Gamma(D_i \rightarrow u_i+W) \approx  2 \Gamma(D_i \rightarrow d_i+Z) 
\approx \frac{1}{16\pi} ( s_L^{d_i} )^2 \frac{M_{D_i}^3 }{v^2}.
\label{eq:BRZ2}
\ee
In view of Section 5.1, the total widths of $U_i$, and similarly for $D_i$, can be written as
\be
\Gamma_{Tot} (U_i)|_{\rm PC}
\approx \frac{3}{32\pi} \left( \frac{m_{u_i} M_U^2 }{v f^u } \right)
\approx 0.1~{\rm MeV} \left(\frac{m_{u_i} }{\rm MeV} \right)\left(\frac{M_U^2}{f^u~{\rm TeV} }
\right)
\ee
or
\be
\Gamma_{Tot} (U_i)|_{\rm PB}
\approx \frac{3}{32\pi} \left( \frac{m_{u_i} M_U^2 M_{U_i} }{v^2 (f^u)^2 } \right)
\approx 0.5~{\rm eV} \left(\frac{m_{u_i} }{\rm MeV}\right)^2 
\left(\frac{M_U^2 M_{U_i}}{(f^u)^2~{\rm TeV} } \right)~.
\ee
Taking the last factor in the r.h.s. of these equations equal to unity, these widths range from about $0.1$ MeV for $U_1$ to about $200$ MeV for $D_3$ in the PC case, whereas they go from a fraction of $1$ eV for $U_1$ to about $1$ MeV for $D_3$ in the PB case.

All this is based on the mass mixings described in Sect. 4. One can ask if these decay properties could be changed by the presence of higher dimensional operators consistent with the symmetries. There is no such operator at any relevant level in the Parity Conserving case. In the Parity Breaking case the operator
\be
\Delta L = \frac{f^u}{\Lambda^3} (\bar{U}_R^i \gamma_\mu u_R^i) \Sigma_j (\bar{q}^j \gamma_\mu q^j), 
\ee
if present, would make the decay of  $U_1$ and $U_2$ dominated by the modes
\be
U_i \rightarrow u_i + \bar{f} f
\ee
where $f$ is any elementary fermion, although still with a small width of about $0.1$ MeV.
Similar considerations hold for the $D_i$.
  
For several aspects (production cross-sections and leading 
decay modes into $W+q$) the phenomenology of these 
heavy states at colliders is quite similar to that of 
sequential fermion families within the SM (see e.g.~\cite{Frampton:1999xi}).
Beside the narrow decay widths, an important difference is the large 
neutral-current branching fraction into $Z$ bosons, as indicated 
in eqs.(\ref{eq:BRZ1})--(\ref{eq:BRZ2}). The $Z+q$ final state,
which can have a non-negligible branching fraction also for sequential quarks 
under specific circumstances~\cite{Frampton:1999xi},
is definitely the most interesting one for searches at hadron colliders. 
According to a recent CDF study~\cite{Aaltonen:2007je}, 
a bound of about $270$~GeV can be set on the mass of 
the partner of the $b$ quark, assuming BR($B \to b Z$) =100\%.

\section{Summary and Conclusions}

A consistent description of 
\begin{itemize}
\item unitarity in $WW$ scattering, 
\item the EWPT, 
\item fermion masses and flavour physics,
\end{itemize}
is greatly eased in the SM by the presence of a Higgs boson, which makes its search, perhaps in a supersymmetric realm, a primary task of the LHC. The competitive view, based on a strong dynamics, has definitely a harder time in achieving the same goals, at least when one tries to come to concrete models.  However we do not forget, on one side, the scanty direct experimentation at the Fermi scale or above it and, on the other side,  that the SM itself is  likely 
to be an effective theory.  Altogether, this still motivates the study of possible generic features of strongly interacting theories of Electroweak Symmetry Breaking  by making the least possible reference to   explicit models. 

Along these lines,
in this work we have analyzed the properties and the constraints on possible composite fermions that might result quite naturally from the strong dynamics. 
As already pointed out, this is made possible at all by focusing on two approximate symmetries:
\begin{itemize}
\item a chiral $SU(2)_L\times SU(2)_R$ that breaks down to its diagonal subgroup,
\item a  symmetry $G_f$ that enlarges the flavour symmetry of the SM, in absence of the Higgs doublet, to include the flavour symmetry of the composite fermions themselves.
\end{itemize}
While the first is a widely accepted feature of strong EWSB, the second one is meaningful and useful if one takes the view that the masses of the standard {\it elementary} fermions only arise from their mixing with the composite fermions. We find this assumption coherent with the picture that  the standard fermions do not participate in the strong dynamics, which is the source of EWSB, whereas their   masses  do break the electroweak symmetry. We hope that the multiplicity of the composite fermions needed to this purpose be explained by the strong dynamics.

Our results can be summarized as follows. We consider Singlets, Doublets and Triplets under the custodial $SU(2)_{L+R}$. In a truly strongly interacting theory of EWSB we find that Doublets and Triplets are faced with difficulties, although a generic analysis like ours cannot exclude them. On the other hand the Singlets, whose composite nature  is admittedly hidden, may nevertheless play an important role. This is in particular the case in flavour physics, where we show that it becomes pretty natural to keep the CKM picture of the SM with MFV. Under this assumption, we can specify the fine structure of the spectrum of the composite fermions, three of charge $2/3$, $U_i$, and three of charge $1/3$, $D_i$, and their decay properties, relevant to the search at the LHC.  Equally specified is their loop contribution  to the $T$-parameter in the EWPT and to several flavour observables. The correlation between these contributions excludes a significant effect of the Singlets in the $T$-parameter. At the same time the non observation of significant deviation in flavour physics from the SM does not set any strict  lower bound on the their  masses. The search for the heavy quarks, with a $30\%$ branching ratio into an ordinary quark and a $Z$-boson, might be feasible up to significantly large values of their masses even in the early stage of the LHC. Whereas we have concentrated on coloured states, all the picture can be naturally extended to leptons. Several aspects of the phenomenology of these composites, if they exist at all, deserve further study.

\section*{Acknowledgements}

We thank Slava Rychkov for useful discussions in the early stage of this work and Paolo Lodone for help in checking the loop calculations. 
R.B. was partially supported by the EU under RTN contract
MRTN-CT-2004-503369 and by MIUR under the contract PRIN-2006022501.
G.I.~was partially supported by the EU
under RTN contract MTRN-CT-2006-035482.

\appendix

\section{Heavy-fermion effects in FCNCs}

Following the notation of Buras~\cite{Buras:1998raa}
we write  the leading electroweak contributions 
to $b\to s$ FCNC transitions in the SM as follows:
\bea
  \cH (\Delta B = 2)  &=& \lambda_t \frac{G^2_{\rm F} m_W^2 }{4\pi^2} 
   ~S_0(x_t)~ 
   (\bar b_L \gamma^\mu s_L)^2
\\
  \cH ( b\to s\nu\bar\nu ) &=& \lambda_t \frac{G_{\rm F}}{\sqrt 2}
   \frac{ 2\alpha}{ \pi \sin^2\Theta_{\rm W}} \lbrack C_0(x_t)-4B_0(x_t)\rbrack 
   (\bar s_L \gamma^\mu d_L) (\bar \nu \gamma_\mu \nu)
\\
  \cH ( b\to s\mu\bar\mu ) &=& \lambda_t \frac{G_{\rm F}}{\sqrt 2}
   \frac{2 \alpha}{ \pi \sin^2\Theta_{\rm W}} \lbrack C_0(x_t)+B_0(x_t)\rbrack
   (\bar s_L \gamma^\mu d_L) (\bar \mu \gamma_\mu \mu)
\\
  \cH ( b\to s\gamma ) &=&  \lambda_t \frac{G_{\rm F}}{\sqrt 2} 
  \frac{e}{8\pi^2} ~D'_0(x_t)~   m_b \bar b_R \sigma^{\mu\nu} F_{\mu\nu} s_L
\\
  \cH ( b\to s G ) &=&  
  \lambda_t \frac{G_{\rm F}}{\sqrt 2} \frac{g_s}{8\pi^2}
   ~E'_0(x_t)~  m_b \bar b_R \sigma^{\mu\nu} T^a G^a_{\mu\nu} s_L
\eea
where $\lambda_t=V^*_{tb}V_{ts}$ and $x_t=m_t^2/m_W^2$. 
The corresponding terms for 
$b\to d$ and  $s\to d$ transitions --in the limit where we can
neglect the charm quark mass-- are obtained by replacement 
of the CKM factor and by $m_b\to m_s$.
The explicit expressions of the loop functions are:
\bea
S_0(x)&=&\frac{4x-11x^2+x^3}{4(1-x)^2}
          -\frac{3x^3 \ln x}{2(1-x)^3}~,
\qquad\qquad\qquad   
S_0(x) ~\stackrel{ x\to \infty}{\longrightarrow}~  \frac{x}{4}
\\
C_0(x)&=& \frac{x}{8}\left[ \frac{x-6}{x-1}
  +\frac{3x+2}{(x-1)^2}\;\ln x\right]~,
\qquad\qquad\qquad\   
C_0(x) ~\stackrel{ x\to \infty}{\longrightarrow}~  \frac{x}{8} 
\\
B_0(x) &=&\frac{1}{4} \left[ \frac{x}{1-x}
 +\frac{x\ln x}{(x-1)^2} \right],
\qquad\qquad\qquad\qquad\ \ 
B_0(x) ~\stackrel{ x\to \infty}{\longrightarrow}~  -\frac{1}{4} 
\eea
\bea
D'_0(x) &=& -\frac{(8x^3 + 5x^2 - 7x)}{12(1-x)^3}+ 
          \frac{x^2(2-3x)}{2(1-x)^4}\ln x~,
\qquad 
D'_0(x) ~\stackrel{ x\to \infty}{\longrightarrow}~  \frac{2}{3}
\\
E'_0(x)&=&-\frac{x(x^2-5x-2)}{4(1-x)^3} 
+ \frac{3}{2} \frac{x^2}{(1 - x)^4} \ln x,
\qquad\quad\ \
E'_0(x) ~\stackrel{ x\to \infty}{\longrightarrow}~  \frac{1}{4}
\eea

\medskip

In the cases where the $x\to \infty$ limit is finite,
the modification of the coefficient functions 
due the addition of the heavy state is
\be
F(x_t) \to c^2 F(x_t) +s^2 F(x_T)= F(x_t) +s^2[ F(x_T)-F(x_t) ], 
\qquad F=B_0,D_0',E_0'
\ee
where $x_T=M_T^2/m_W^2$. As expected, the correction 
vanishes for $s=0$ or $x_t=x_T$ (which corresponds 
to setting $m_L=m_R$ and $M_Q=0$ in the mass matrix). 
In this specific case the correction 
vanishes also for $M_T,m_t \gg m_W$.
Note, however, that the $x_t\to \infty$ limit is not 
necessarily a good numerical approximation for the 
physical top-quark mass. For instance in the 
$b\to s\gamma$ case $[D_0'(\infty)-D_0'(x^{\rm phys}_t)]\approx 0.7 D_0'(x^{\rm phys}_t)$.

For the amplitudes where the $x\to \infty$ limit is 
not finite ($\Delta F=2$ box and $Z$ penguin) 
we cannot express the result using only SM loop functions. 
The result can be written in the following general form: 
\bea
 F(x_t) &\to&
\left[ c^4 \bar F (x_t,x_t) +s^4 \bar F (x_T,x_T) + 2c^2s^2 
\bar F(x_t,x_T) 
\right]  - c^2 \Delta F (x_t) -s^2 \Delta F (x_T) \no \\
&& = F(x_t) + s^2 
\left[ (s^2-2) \bar F (x_t,x_t) + s^2 \bar F (x_T,x_T) + 2c^2 
\bar F (x_t,x_T) 
\right] \no \\
&&  - s^2 [ \Delta F_S (x_T) - \Delta F_S (x_t) ]~, 
\label{eq:FXTgeneral}
\eea
where 
\be
\Delta F(x)\equiv \bar F(x,x)-F(x)~, \qquad\quad   
\frac{\Delta_F(x)}{F(x)} ~\stackrel{ x\to \infty}{\longrightarrow}~0~,
\qquad\qquad F=S_0,C_0~.
\ee
As expected, also in this case the correction vanishes
for $s=0$ or $x_t=x_T$. 
The leading functions $\bar F(x,x)$ for 
the two relevant cases can be identified by 
the explicit calculation of the  diagrams with two
different heavy propagators. In the limit $m_W \to 0$
the result is particularly simple and universal,
\be
\frac{\bar F(x_T,x_T) }{ \bar F(x_t,x_t) } = \frac{M_T^2}{m^2_t}~, 
\qquad \frac{\bar F(x_T,x_t) }{ \bar F(x_t,x_t) } = 
\frac{M_T^2}{M_T^2 -m_t^2}\ln \left(\frac{m_T^2}{m_t^2}\right)~,
\ee
leading to the function $R(x_t,x_T)$ in Eq.~(\ref{eq:Runiv}).
The reason for this universality can be understood by the 
fact that in the gaugeless limit the only UV and IR finite 
integral with two heavy propagators, which exhibit the right 
grow for $m_1=m_2\to\infty$, is 
\be
 \int dl^2 \frac{ m^2_1 m^2_2 }{(l^2+m^2_1)(l^2+m^2_2)}
=  \frac{ m_1^2 m_2^2 }{m_1^2 -m_2^2}  \ln\left( \frac{m_1^2}{m_2^2} \right)
\ee

The complete expressions of the loop functions,
necessary to evaluate also the subleading terms, are 
\bea
\bar S_0(x_1,x_2) &=& -\frac{3 x_1 x_2 }{4 (1-x_1) (1-x_2)}
+ \left[~\frac{ x_1 x_2(4-8 x_1+x_1^2)}{4 (1-x_1)^2 (x_1-x_2)} \ln(x_1)
~+~ [x_1 \leftrightarrow x_2]~ \right], \qquad \\
\Delta S_0(x) &=& 0~, 
\eea
for the box diagram, and 
\bea
\bar C_0(x_1,x_2) &=&
 \frac{ x_1^2 (x_2-1)}{8 (x_1-x_2)(x_1-1)} \ln(x_1) 
~+~ [x_1 \leftrightarrow x_2]~,\\
\Delta C_0(x) &=& -\frac{5}{8} \frac{x (1-x+x \ln(x))}{(x-1)^2}~,
\eea
for the $Z$ penguin.

The general decomposition in (\ref{eq:FXTgeneral})
applies also the the flavour-conserving 
non-decoupling effects in $\delta T$ and $Z\to \bar b b$.
Since the approximate expression in  Eq.~(\ref{eq:Runiv})
turns out to be an excellent numerical approximation for 
the $Z \to \bar b s$ penguin, we have used it 
in Fig.~\ref{fig:bounds} to estimate 
the bound derived from $Z\to \bar b b$.

\end{document}